# Unusual electronic properties of sub-nanosized magnesium clusters

Stanislav K. Ignatov, [1*] Artëm E. Masunov[2]

[1] *Lobachevsky State University of Nizhny Novgorod, Nizhny Novgorod 603950, Russia*

[2] *NanoScience Technology Center, University of Central Florida, Orlando, Florida 32826, USA*

* skignatov@gmail.com

ABSTRACT

The electronic parameters and, in particular, the isotropic electrostatic polarizability (IEP) of sub-nanoscale magnesium clusters were studied in an expanded set of 1237 structurally unique isomers found in the course of direct global DFT optimization of the structure of $Mg_2$-$Mg_{32}$ clusters at the BP86/6-31G(d) level, as well as using global optimization based on DFT-calibrated MTP potential for some larger structures. The calculation of the polarizability at the same DFT level reveals an unusual property of the IEP – the dependence of the IEP of the most favorable isomers on the cluster nuclearity *n* is linear with a high correlation coefficient, and its value for each *n* is close to the minimum value among all found isomers of a given nuclearity. These features take place independently on the cluster structure which allows hypothesizing that the energetic favorability of a cluster structure is connected to their polarizability.  A possible explanation of the observed dependence, its significance for quantum chemistry, and the possibility of practical application are discussed.

## Introduction

Sub-nanosized metal clusters are polyatomic compounds of variable nuclearity (the number *n* of atoms in a cluster), occupying an intermediate position between atoms and nanoparticles, differ from the latter in a lower degree of crystalline ordering and in greater structural diversity. Their typical diameter is of 0.2–2 nm which corresponds to the nuclearity up to 150-200 atoms. When the size increases, the large crystalline domains appear in a particle which makes them similar to polycrystals, and the standard theoretical methods for the description crystalline structures can be applied to evaluation of various physic-chemical properties (see, e.g., study of [1] where adsorption energy on Pt clusters went to the limiting values of crystal surface at $n > 147$). In contrast, the cluster structures at lower *n* (typically $n$=2-150) are poorly predictable, both at the levels of simple "chemical intuition" and standard metallic potentials which are used with high success to the description of bulk metal properties or the properties of metal surface. Among quantum chemical methods, only DFT methods can usually be used for the effective search for favorable structures of small clusters, although one cannot always be sure of the reliability of their predictions, since the benchmark high-level theories can often be applied in a limited range of nuclearities only. The same applies to the so-called novel "potentials of quantum accuracy" (GAP,[2] SNAP,[3] MTP,[4] ACE [5]), since they are usually calibrated by DFT results in a limited range of nuclearities. In this situation, the most reliable method for the sub-nano cluster structure prediction remains the direct DFT global optimization. At the same time, the properties of sub-nanosized metallic particles, both mono- and polyatomic ones, are of great interest because they frequently manifest higher activity [6,7] and selectivity [8] in catalysis [9-11], can serve as the base elements for the modern and further nanoelectronic,[12,13] or spintronic[14,15] devices or as a base for the novel nanodevices manifesting e.g., neuromorphic properties[16]. Previously, using the example of magnesium clusters, we demonstrated that the number of the cluster isomers which can be formed for the given *n*, although is high and quickly grows with *n*, nevertheless remains much lesser than the number of mathematically predicted number of connected graphs of the same number of vertices [17]. This fact allows to explore the complete set of isomers at once (at least for some *n*), trying to search for the individual representatives with useful properties. In a recent study [18], we carried out such a search in the extended set of isomers of clusters $Mg_2$-$Mg_{32}$ comprising 1237 isomeric structures located in the direct DFT global optimization. The magnesium clusters are the convenient object for such studies because they are simple for DFT calculations, their ground state is a singlet spin state, and many reference data are known for them, both experimental and theoretical. In the study [18], we considered mostly energetic and structural properties, which allowed us to locate the most stable representatives, the energy distributions for a complete set of isomers as well as the distributions for separate nuclearities. The analysis of these distributions allowed to predict that the size- and shape-selected cluster formation can be performed for some *n*, in particular, for *n*=10 and 20. Also, some new tubular structures and the new dependences between the structural features of clusters were established. In the current study, we continue the investigation of the extended set of magnesium clusters, focusing now on their electronic properties. Among these electronic properties, we find that the extended set of isomers demonstrates quite unusual feature of their isotropic electrostatic polarizability (IEP): IEP of the most favorable structures (and of only them) is linearly dependent on *n*, and, even more surprising, these linearly dependent values are close in their magnitude to the minimum IEP values for the set of isomers with the given *n*. This property is rather unusual because the most favorable isomers typically have quite different geometry structure. Moreover, this property, if it will be common for other metals, can be extremely useful since it gives the mean for the prediction how the random cluster structure is close to the global minimum – the mean which opens new ways in cluster structure prediction. We also show that the connection between the polarizability and the DFT energy of cluster cannot be simply derived from any known models of polarizability although some models reproduce its linear dependence (but do not "explain" it directly). The paper is organized as follows. In the next section, we briefly describe the methodology of cluster generation and evaluation of their properties. In the beginning of the Results and Discussion section, we describe the calculated cluster polarizability and the found dependencies of it on different structural properties. We also discuss here some principles and possible explanations of the found dependencies including the applications of some known models of molecular polarizability. The Concluding remarks section contains a discussion of possible implications of the found regularities and the directions that could be explored in order to rationalize the patterns found.

## Calculation details

*Set of isomeric structures.* Structures of magnesium clusters $Mg_2$-$Mg_{32}$ used in the analysis were generated in the course of the direct DFT global optimization combined with graph generation algorithm and manual construction of the highly symmetric structures as was described in detail in refs [17,18]. In brief, the initial structures were generated from a complete set of connected graphs with *n* vertices as well as using the evolutionary algorithm combined with *taboo*-search to avoid the generation of similar structures. The geometries of generated structures were thoroughly optimized at the BP86/6-31G(d) DFT level with higher optimization criteria (*Tight* stopping criteria and *UltraFine* DFT grid implemented in Gausian16 [19] program) and unique final structures were selected on the basis of two different algorithms of similarity evaluation in order to establish the unique isomers. Such a procedure continued until new unique structures ceased to appear. The Cartesian coordinates of 543 structures of $Mg_2$-$Mg_{13}$ located by this method were reported earlier in Supporting Information for ref [17]. In a course of this work, about 9000 optimizations were carried out and about 820000 points of potential energy surface (PES) were explored in total. All the located unique structures are the true local minima of PES as was proven by the frequency calculations at the same theory level. The remaining 694 structures of $Mg_{14}$-$Mg_{32}$ were located by the similar methodology, although without achieving the limit of all possible structures, with the DFT optimization using normal optimization criteria as was described in [18]. Their Cartesian coordinates were published as Supplementary Information for ref [18].

*DFT calculations.* All energies, polarizabilities and other electronic properties were calculated at the BP86/6-31G(d) level of DFT theory which was proven to be the best level of theory describing the results of combined CCSD//MP2/cc-pVTZ level of theory for $Mg_2$-$Mg_7$ clusters.[20,21] Our previous evaluation of the performance of this DFT level for $Mg_{10}$ cluster shows that this theory level agrees well with B3PW91/6-31G(d) results whereas PBE0/6-31G(d) results are somewhat worse.[17] Also, comparison of BP86 results with the results of CCSD(T)/cc-pVQZ and MP2/cc-pVQZ shows that the bond lengths in small clusters $Mg_2$-$Mg_4$ are remarkable underestimated at the DFT level but they going better as nuclearity increases. At the same time, this DFT level better reproduces the cluster energies of CCSD(T) than it takes place in the case of MP2. All calculations were carried out using the Gaussian16 [19] program.

*Calculation of electronic properties.* For the complete set of 1237 structures of $Mg_2$-$Mg_{32}$ obtained as described above, the additional single point calculation was carried out at the BP86/6-31G(d) theory level with *UltraFine* DFT grid in order to calculate the MO vectors of Kohn-Sham orbitals, HOMO/LUMO energies, the electrostatic polarizabilities, and other electronic properties. The obtained MO vectors were then processed with ChargeMol program [22,23] in order to calculate the Density Derived Electrostatic and Chemical (DDEC6) charges, bond orders, and the atom valences. [23,24] Other electronic characteristics (Coulson's, Mayer's, Wiberg's bond orders, charges and valences) were extracted from the calculation log files or calculated on the basis of MO vectors. Additionally, the values of coordination numbers, HOMO and LUMO energies, cluster energies, and, in some cases, NBO parameters were extracted and analyzed. DDEC6 method used here is the relatively new method for the bond order evaluation [23,24] which has significant advantages before the "classical" bond order descriptions including the Coulson's, Mayer's, Wiberg's, and NBO bond orders and charges. Namely, we found that it gives much more clear description of bonds between the Mg atoms. For example, the above-mentioned classic bond order evaluation schemes find the large number of bonds inside the clusters (virtually for all pairs of atoms) with close values of bond orders. At the same time, the DDEC6 bond orders are much more differentiated between near-by and far atoms which makes this method more convenient for analysis of the chemical bonding inside the metal clusters.

*Cluster energies.* In the following, we use the per atom binding energy of cluster (also termed as a reduced cluster energy) defined as

$$E_b(n,m) = E_{tot}(n,m)/n - E_{tot}(1,1)$$

Here, $E_{tot}(n,m)$ is the total energy of the optimized cluster structure for the *m*-th isomer of cluster of *n* atoms, i.e. the DFT-calculated sum of its electronic energy and the energy of nuclear repulsion. $E_{tot}(1,1)$ is the total energy of a single Mg atom (–200.0697059 Hartree at the BP86/6-31G(d) with *UltraFine* grid). With this definition, all $E_b(n,m)$ values are negative, and their dependence on *n* and *m* was reported in Fig.1 of ref [18]. For the given *n*, all *M* isomers are ordered in an ascending order of their binding energies $E_b(n,m)$ ($m$=1,…,*M*). Thus, $E_b(n,1)$ is the binding energy of the most favorable isomer of $Mg_n$ with the lowest $E_b$ value (maximum in its absolute value) among *M* isomers of the given *n*. Note, that, despite energy ordering within individual *n*, the binding energies of clusters with different *n* can be greater or less than each other. We also analyze the relative energy of cluster within the given *n* defined as

$$E_{rel}(n,m) = E_b(n,m) - E_b(n,1),$$

i.e. the energy difference between the binding energy of *m*-th isomer and the corresponding most favorable isomer of the same nuclearity.

*Electrostatic polarizability* of an isolated atom is defined as a scalar coefficient α between the induced dipole moment **p** of an atom and the external weak electrostatic field inducing this moment: $\mathbf{p} = \alpha \mathbf{E}$. This simplified finite field formula differs from the strict definition in terms of the field derivatives since we do not discuss here the nonlinear effects, which are not relevant to the purposes of the present study. In the case of polyatomic structure, the polarizability is a second rank tensor **α** with components

$$\boldsymbol{\alpha} = \begin{pmatrix} \alpha_{XX} & \alpha_{XY} & \alpha_{XZ} \\ \alpha_{YX} & \alpha_{YY} & \alpha_{YZ} \\ \alpha_{ZX} & \alpha_{ZY} & \alpha_{ZZ} \end{pmatrix}$$

For optically inactive molecules, the tensor is symmetric. To avoid the dependence on coordinate framework, two quantities are usually introduced, isotropic polarizability $\alpha_{iso}$ and polarizability anisotropy $\alpha_{aniso}$:

$$\alpha_{iso} = \frac{1}{3}\left(\alpha_{XX} + \alpha_{YY} + \alpha_{ZZ}\right)$$

$$\alpha_{aniso} = \left[\frac{(\alpha_{XX} - \alpha_{YY})^2 + (\alpha_{YY} - \alpha_{ZZ})^2 + (\alpha_{ZZ} - \alpha_{XX})^2 + 6(\alpha_{XZ}^2 + \alpha_{XY}^2 + \alpha_{YZ}^2)}{2}\right]^{1/2}$$

Polarizability tensor can be directly calculated within the DFT method for a molecular structure using many quantum-chemical programs. In these calculations, only the purely electronic polarizability is considered, without any contributions of orientational or ionic terms (no displacements of nuclei from their initial positions due to an external field are allowed).

## Results and Discussion

*Dependence of cluster polarizability on n and m*. Fig.1*a* shows the dependence on *n* and *m* of the calculated isotropic polarizability $\alpha_{iso}$ for 543 isomers of $Mg_2$-$Mg_{13}$ which is the complete set of isomers found for these clusters by the method of the long-lasting structure generation and optimization. The black triangles mark the most favorable isomers, blue triangles – least favorable isomers of each nuclearity. Each point shows the data for an individual isomer, and the color scale indicates $E_{rel}$ in kcal/mol.

(*a*)

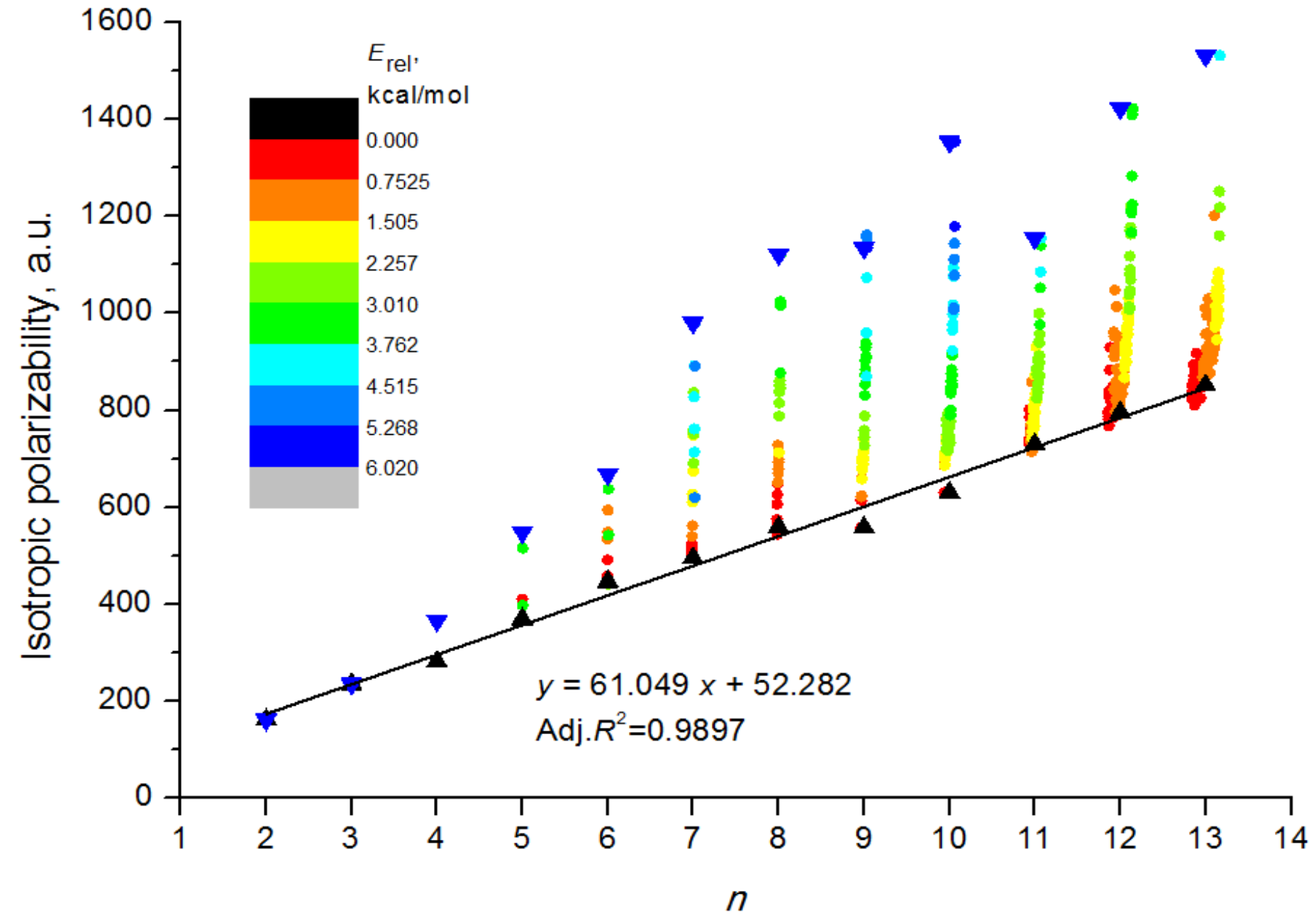

(*b*)

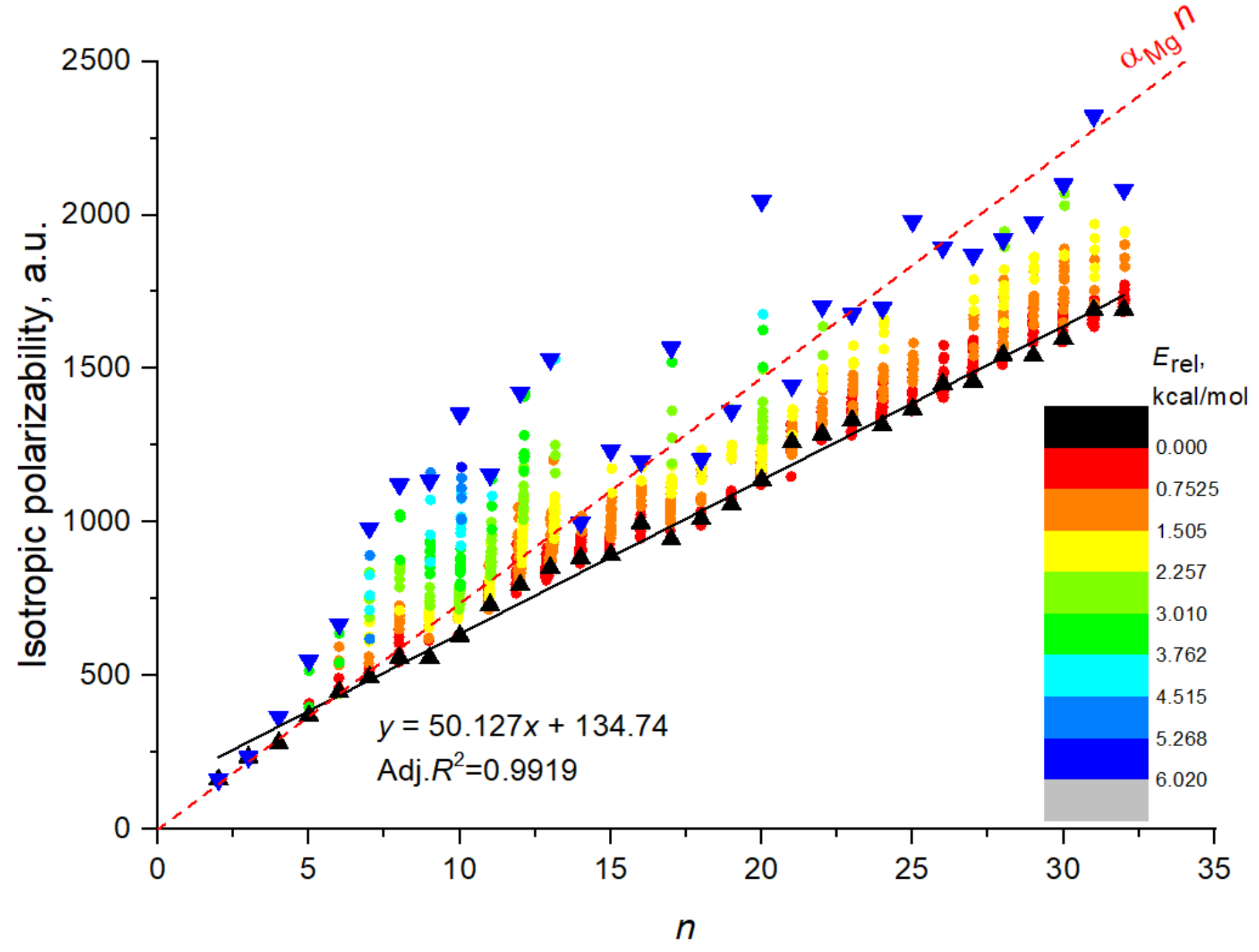

Figure 1. The DFT calculated IEP of magnesium clusters: (*a*) for 543 isomers of $Mg_2$-$Mg_{13}$; (*b*) for 1237 isomers of $Mg_2$-$Mg_{32}$. Black triangles are the most energetically favorable isomers, blue triangles – the least favorable isomers. Color scale designates the cluster energy relatively to the most favorable isomer of the given *n*. Red dashed line – the idealized polarizability of *n* non-interacting Mg atoms. The data points within each individual *n* are slightly shifted in horizontal direction to avoid overlapping.

Striking feature of the presented plot is the linear dependence of $\alpha_{iso}$ on $n$ for the most favorable isomers. Moreover, the polarizabilities of the most favorable isomers are always close to the minimum value of $\alpha_{iso}$ among all isomers for the given $n$. Although $\alpha_{iso}(n,1)$ is the exact minimum value of polarizability not for all $n$, and some isomers with low $E_{rel}$ (red and orange colored) have the slightly lower values in the case of $n$=11-13, the polarizability of the most favorable isomers has a clear tendency to be close to a minimum value. The polarizability of the most favorable isomers is perfectly described by the linear function with high determination coefficient of 0.9897, the equation is shown on the plot. Thus, we note two clear features of the $\alpha_{iso}$ on $n$ and $m$: (1) linearity of $\alpha_{iso}(n,1)$ on $n$; and (2) minimality of $\alpha_{iso}(n,1)$ among all $m$.

The same tendency is also clearly seen for the expanded set of 1237 cluster isomers of $Mg_2$-$Mg_{32}$, see Fig.1*b*. In this case, the linearity is characterized by even higher determination coefficient of 0.9919, although the deviations of some points from linear function are remarkable. The minimality of $\alpha_{iso}(n,1)$ is also not so perfect as for $Mg_2$-$Mg_{13}$, and numerous deviations on this rule is noticeable. Nevertheless, both above tendencies are clearly manifested for the expanded set of isomers as well. It should also be noted that both plots demonstrate the clear (although not perfect) tendency of the isomers with $m$>1 to have the increased polarizability depending on their relative energies (isomers with higher $E_{rel}$ tend to have higher polarizability) as is seen on the base of datapoint color changes in Fig.1. The additional statistical parameters of the linear regression model describing the complete set of isomers are given in Table 1, the column “DFT results”.

Table 1. Statistical parameters of regression formula (1) describing the DFT calculated values of $\alpha_{iso}(n,1)$ ($n$=2-32) and the fitted models estimates for $\alpha_{iso}$ obtained in the induced dipole approximations.

| Model parameters[a] | DFT results | Induced dipole models X AM | TL | TE | RP | GD |
|---|---|---|---|---|---|---|
| **Fitting results of $\alpha_{iso}(n,m)^X$ against $\alpha_{iso}(n,m)^{DFT}$** | | | | | | |
| Adjusted $a_{smear}$ | – | – | 6.3517 | 1.6366 | 1.0910 | 1.0416 |
| Final RMSD | – | 56267. | 138.06 | 140.44 | 43.45 | 48.86 |
| Negative $\alpha_{iso}$ | 0 | 263 | 0 | 0 | 0 | 0 |
| **Parameters of linear regression (1) for $\alpha_{iso}(n,1)$ estimated with DFT and fitted models** | | | | | | |
| Slope of (1) | 50.12 | 123.13 | 55.75 | 56.40 | 50.16 | 49.44 |
| Intercept of (1) | 134.74 | 210.46 | 95.36 | 82.01 | 104.37 | 111.51 |
| $S_a$ | 0.83 | 24.52 | 0.55 | 0.42 | 0.57 | 0.65 |
| $S_b$ | 15.85 | 471.09 | 10.60 | 8.14 | 10.94 | 12.53 |
| $R^2$ | 0.9922 | 0.4650 | 0.9972 | 0.9984 | 0.9963 | 0.9950 |
| Adj.$R^2$ | 0.9919 | 0.4466 | 0.9971 | 0.9983 | 0.9962 | 0.9948 |
| $SSE \cdot 10^{-4}$ | 4.90 | 4325.52 | 2.19 | 1.29 | 2.33 | 3.06 |
| $SE$ | 41.10 | 1221.29 | 27.48 | 21.09 | 28.35 | 32.50 |

[a] – Description: $a_{smear}$ – smearing parameter value adjusted for the best fit between DFT and models with smearing; *RMSD* – Root means squared deviations between $\alpha_{iso}(n,m)^X$ and $\alpha_{iso}(n,m)^{DFT}$ after fitting for all 1237 isomers of $Mg_2$-$Mg_{32}$; Negative $\alpha_{iso}$ – number of negative $\alpha_{iso}$ (non-physical results) among all 1237 isomers; $S_a$, $S_b$ – standard deviations for slope and intercept in (1); $R^2$, $Adj.R^2$ – determination coefficients; *SSE* – sum of squared deviations of $\alpha_{iso}(n,1)$ from (1); *SE* – standard deviation of regression model $(SSE/(N–2))^{1/2}$.

The red dashed line on Fig.1*b* shows the "idealized" polarizability of non-interacting atoms, which is the simple sum of the individual atomic polarizability, i.e. $n\alpha$, where $\alpha=\alpha_{iso}(1,1)$ is the polarizability of single Mg atom, which is of 73.5 a.u. at BP86/6-31G(d) level of DFT. It is seen that this polarizability is higher than $\alpha_{iso}(n,1)$ for $n>6$ and practically coincide with it for $n=2$-6.

Having $\alpha_{iso}(n,1)$ is linear relatively to the cluster nuclearity,

$$\alpha_{iso}(n,1) = an + b\,, \tag{1}$$

it is interesting to investigate the similar trends for the remaining isomers. Fig.2*a* shows the dependence of the "remaining part" of $\alpha_{iso}(n,m)$ after subtracting the linear function (1) from it, $\alpha_{iso}(n,m)-(an+b)$ , on relative energies for the 543 isomers of $Mg_2$-$Mg_{13}$. As is seen from the plot, the remaining part of IEP demonstrates approximate linear dependence on $E_{rel}$ although the linearity is not perfect (regression expression and its determination coefficients are shown on the plot). The plot also shows that there are no any isomers having $\alpha_{iso}$ lower than the linear function (1) except the most favorable isomers themselves. The similar trends persist in the case of the expanded set of studied isomers presented on Fig.2*b*. It is characterized by the similar linear function with approximately the same determination coefficient of 0.84. However, there are some isomers with IEP laying lower the linear function (lower the dashed line in Fig.2b). There are 144 of such values, about 11.6 % of all 1237 isomers, and most of them appears for $n=26$ (18 of 37 structures), $n=31$ (27 of 40 structures), and for $n=32$ (25 of 35 structures). As is seen from Fig2*b*, all these "improper" structures are located in rather narrow region of relative energies, typically within 0.5 kcal/mol above the most favorable structure. This supports the above-mentioned tendency that the minimum IEP correlates well with the low cluster energy.

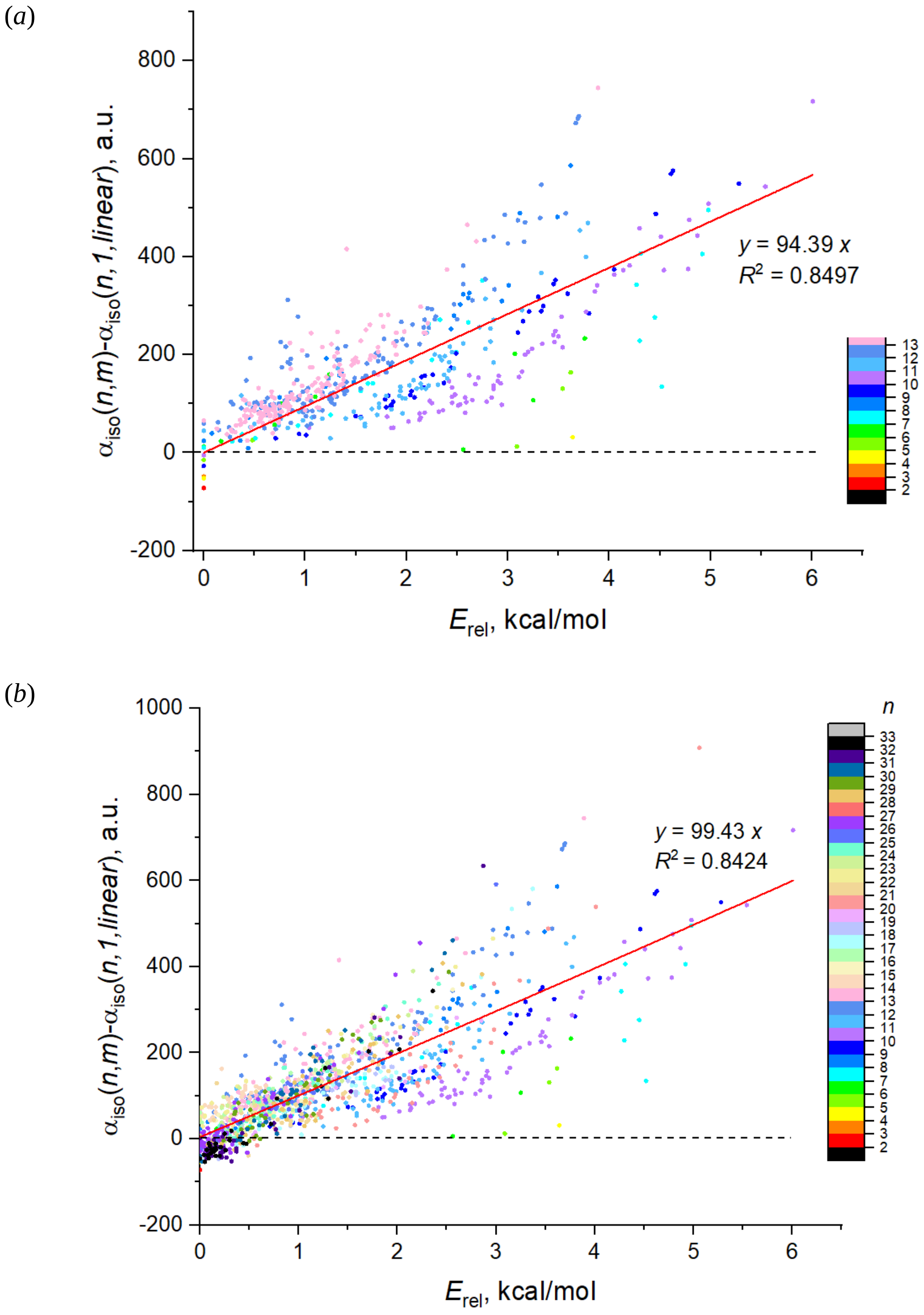

Figure 2. Dependence of the IEP deviation from linear function (1) $\alpha_{iso}(n,m)-(an+b)$ on relative energies of cluster isomers. (*a*) for 543 isomers of $Mg_2$-$Mg_{13}$; (*b*) for 1237 isomers of $Mg_2$-$Mg_{32}$. Black triangles are the most energetically favorable isomers, blue triangles – the least favorable isomers. Color scale designates the cluster nuclearity *n*.

Combining expressions on Figures 1 and 2, one can derive the more general dependence for the IEP as

$$\alpha_{iso}(n,m) = an + b + cE_{rel}(n,m), \quad (2)$$

with $a$=50.13 a.u., $b$=134.74 a.u., and $c$=94.43 a.u./(kcal/mol) for $n$ in the range 2-32, although the confidence level for this regression model is obviously lower than it takes place for the $\alpha_{iso}(n,1)$ in formula (1).

It is also instructive to consider the deviation of the calculated IEP from the "ideal" IEP values for non-interacting atoms $n\alpha$. In Fig.3, the ratio $f=\alpha_{iso}(n,1)/(n\alpha)$ (frequently referred as an enhancement factor) on the relative energy of isomers $E_{rel}(n,m)$ is shown for $n$=2-32, the color scale designates $n$. On this plot, the leftmost data points (with $E_{rel}$=0) correspond to the most favorable isomers, the remaining points demonstrates the dependence of polarizability on $m$. Although it is hard to carry out the exact regression analysis for all points, it is seen that this dependence is close to quadratic $f \sim E_{rel}^2$, with clear dependence of $f$ on $n$.

In contrast with IEP, no similar tendencies were found for the polarizability anisotropy $\alpha_{aniso}$.

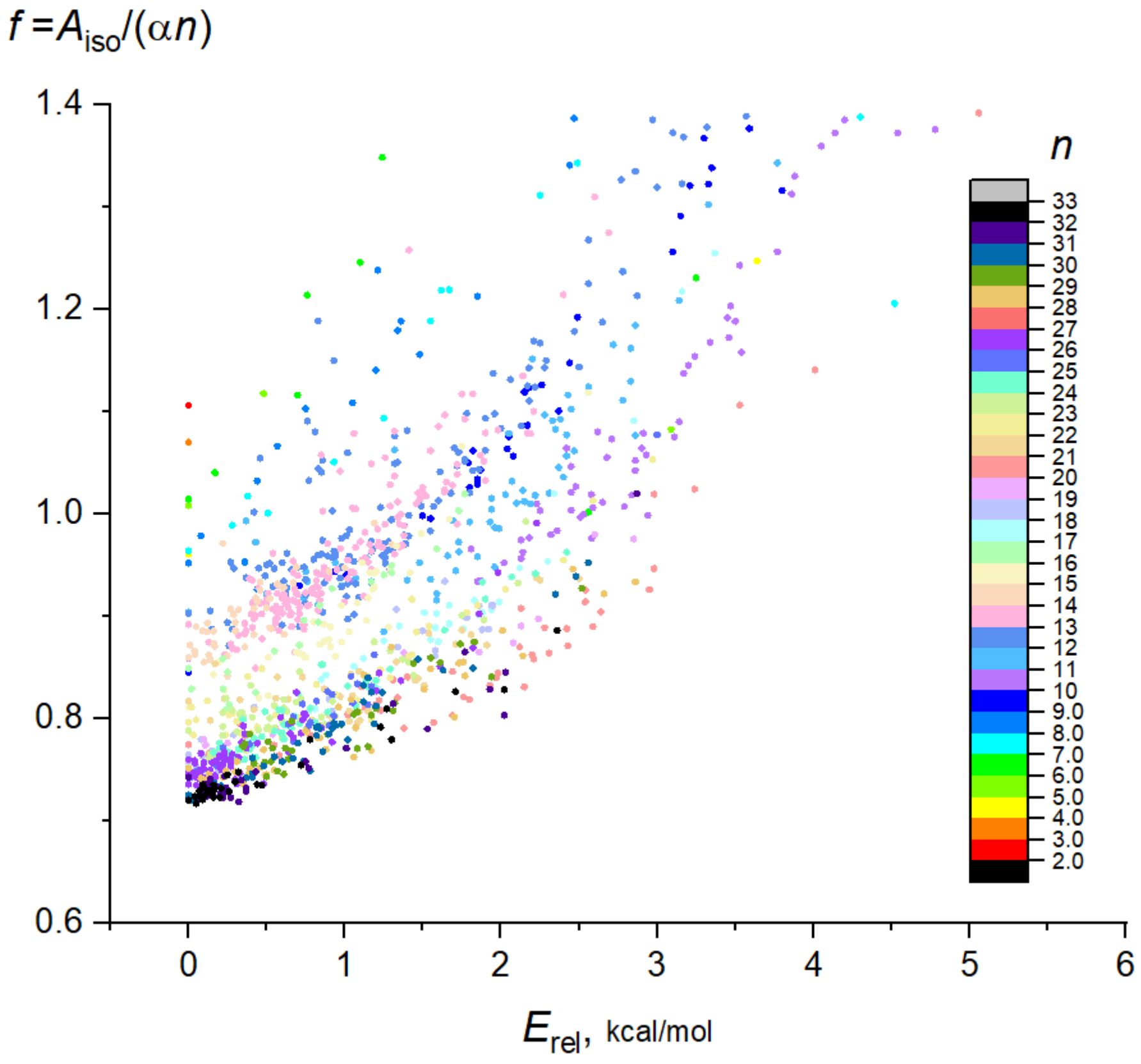

Figure 3. Dependence of enhancement factor *f* on the isomer relative energy for 1237 isomers of $Mg_2$-$Mg_{32}$. Color scale designates the cluster nuclearity *n*.

*Principle of maximum tightness.* Two established features of $\alpha_{iso}(n,1)$ (linearity and minimality) are rather surprising taking into account that the isomers structures are quite different in their geometry. For example, the most favorable isomers of $Mg_2$-$Mg_{32}$ are shown in Fig.4, and the isomers *m*=1–48 of $Mg_{10}$ are shown in Fig.5. As is seen from the Figures, the cluster structures have quite different structural elements including tetrahedral, bipyramidal, decahedral, tubular motives, or highly amorphized fragments. They are quite diverse both among isomers with different *n* and different *m*. Thus, the observed features of polarizability unite the clusters of different structure and energy which allows hypothesizing that some common principle controls the formation of cluster structures. Namely, the structure of a cluster is determined by such a mutual adjustment of the electronic and nuclear subsystems that ensures the maximum stability of the electron shell of the molecule with respect to external polarization. In other words, in order to ensure the most favorable nuclear configuration of the cluster, the nuclei of atoms occupy such positions as to ensure the maximum "rigidity" (“tightness” or "compactness") of the electron shell.

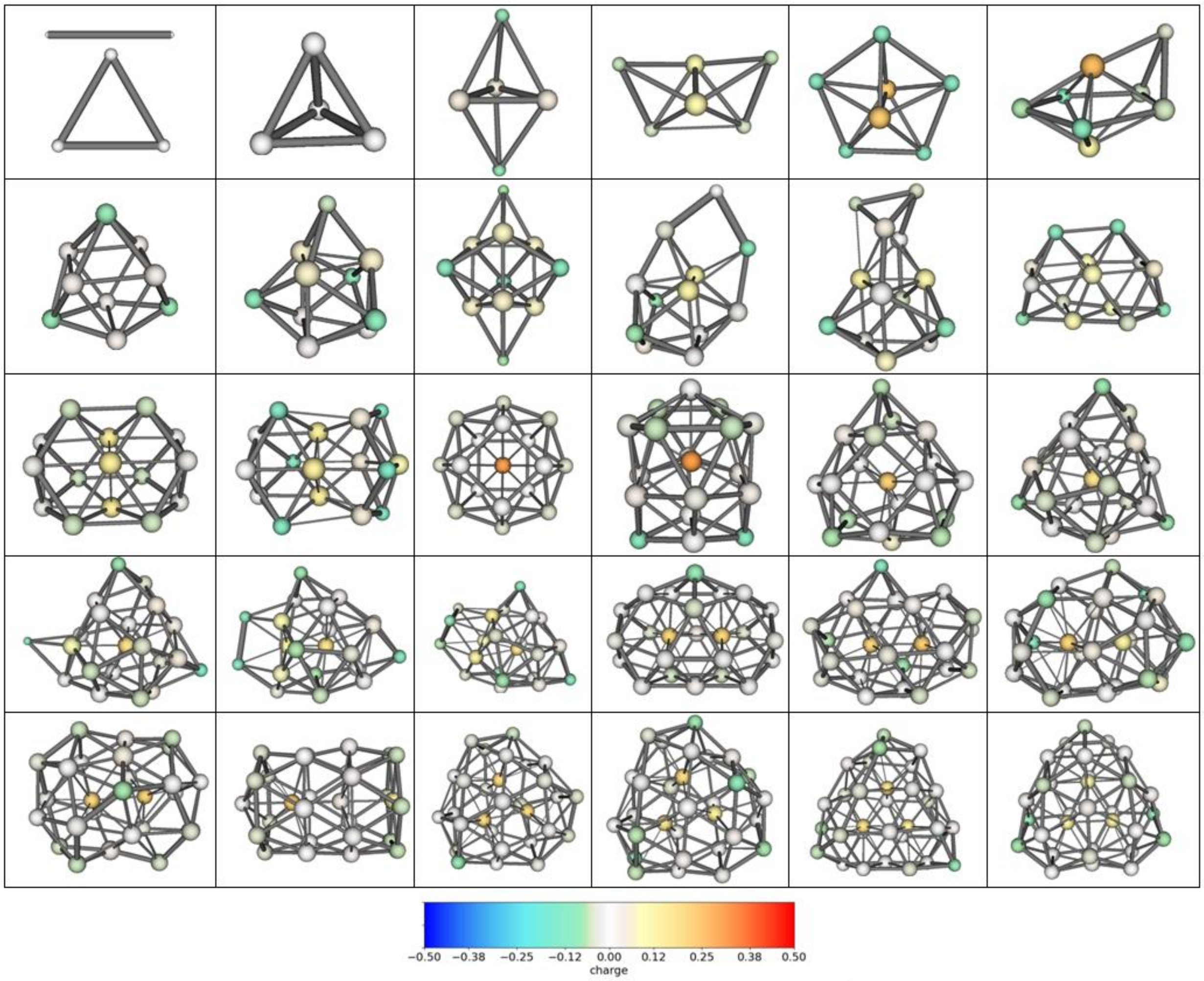

Figure 4. Structures of the most energetically favorable isomers $Mg_2$-$Mg_{32}$. Color scale designates the DDEC6 atomic charges. The bar thickness designates the DDEC6 bond order (the values higher 0.1 are shown).

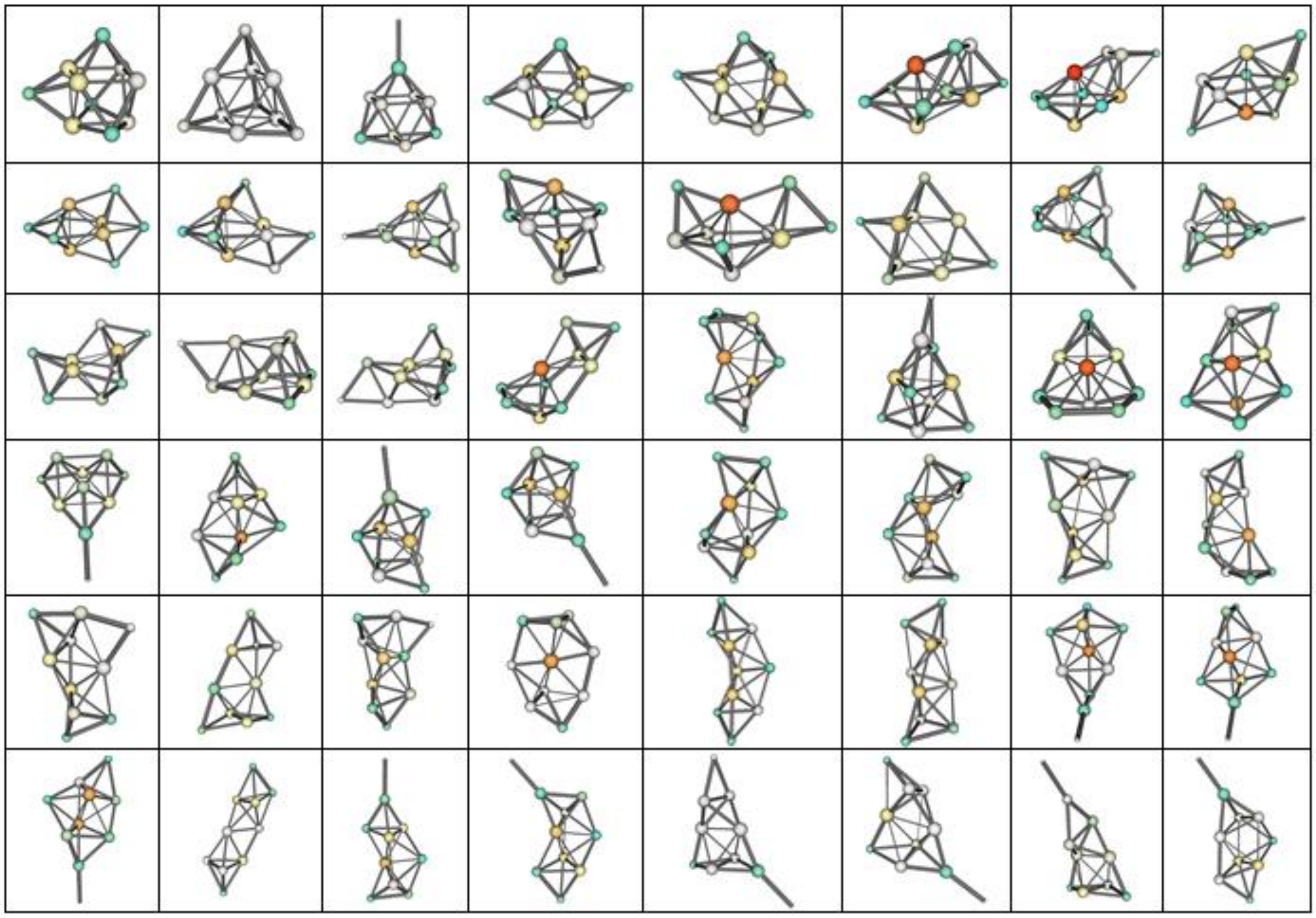

Figure 5. 48 low-lying isomers of $Mg_{10}$. *m* increases from left to right, by rows from up to down, the most favorable isomer is the upper-left one.

The foregoing allows us to formulate (at least for sub-nanoscale magnesium clusters) a "*principle of maximum tightness of electronic shell*" (*PMT*): the most energetically favorable cluster structure has the electronic shell, polarizability of which is close to the minimum and depends linearly on the number of atoms.

This principle is not absolutely strict, as is seen from the deviations of individual polarizability from the linear function, and also from the non-perfect minimality of $\alpha_{iso}(n,1)$ in the case of some *n*. However, the repeatability of the observed trends for broad range of *n* cannot be occasional. It is also interesting question for further studies, whether these trends will take place for other metals and how common this principle for polyelemental clusters and molecules.

The formulated principle has very important implications. First, it allows one to specify the most favorable isomers among arbitrary structures without calculating the energy, only on the basis of their polarizabilities. This is not so important for quantum-chemical calculations, because the computational costs for the energy calculation are much lower than for the calculation of polarizability. However, if one has any simplified scheme for the polarizability estimation, this can be important in a practice. Now, such simplified approximate calculations of polarizabilities are broadly used for the development of polarizable force fields for molecular dynamics (see e.g. AMOEBA+[25,26] and many other polarizable force fields). Thus, if there is a way to quickly estimate $\alpha_{iso}$, one can search for the global minimum without a quantum chemical calculation at all, at least obtaining the good initial structures for the further DFT optimization. Second, it is extremely useful that this principle allows us to specify whether the given structure is close to the global minimum. The problem of modern global optimization algorithms is that the location of a new favorable structure does not guarantee that the global minimum is located and there are no any rules indicating that we should stop the explorations of structures. On the basis of the formulated principle, we can feel certain that if $\alpha_{iso}$ of a cluster is close to the known linear dependence on *n*, then this structure is close to the global minimum.

*Predictive power of PMT.* The assumptions made above on the possibility of using PMT can be directly verified on the basis of test calculations. Here, we present the preliminary results for only single example; the thorough examination will be given elsewhere. To test the predictive ability, we made a search for the global PES minima of the $Mg_{35}$ cluster using the method of effective interatomic potential MTP [4,27] trained on the cluster structures of $Mg_{25}$-$Mg_{32}$ (taken from previous DFT calculations). All calculations with MTP potential were carried out with the MLIP-2 program,[28] the trained MTP potential parameters (file in the format of MLIP-2) can be obtained from authors on demand. The structure of global minimum for the cluster $Mg_{35}$ located with the trained MTP potential is shown in Fig. A1 of Appendix. On the basis of formula (1) and the corresponding statistical parameters of regression from Table 2, we predict that $\alpha_{iso}(35,1)$ has the value of $(1889 \pm 41)$ a.e. with the confidence interval indicated as the regression standard deviation of $\alpha_{iso}(n,1)$ (*SE* in Table 1, column “DFT results”). Using the trained MTP potential, we optimized 10000 random structures of $Mg_{35}$, and carried out the DFT calculations of IEP for the most favorable structure. Its $\alpha_{iso}$ is equal to 1857 a.u., only 32 a.u. lower than the predicted value, which is within the above confidence interval. Although the test performed for only single nuclearity is not reliable enough, and further analysis should be carried out in further studies, the obtained results demonstrate that the predicted values is not restricted by the ranges of *n* studied here and can be used for the extrapolation.

*Possible explanations for established features.* The above unusual features of IEP raise a question on the possible origin and explanation of these regularities. It should be noted that some dependences of electrostatic polarizabilities of the nanoparticles (NPs) of increasing size is known in the field of nanoscience. Namely, Kim et al.[29-31] reported a thorough analysis of NPs polarizability ($\alpha_{NP}$) and the enhancement factor $f=\alpha_{NP}/(n\alpha)$ for model NPs of different shapes when their size grows (typically up to $10^3$-$10^4$ “atoms” or “molecules” forming the NP). It was established that in some ranges of *n*, the *f* values slowly grow and converges to a limiting value. In the region of small *n* and for some simple shapes of NPs, this growth can be considered as an approximate linear. It was also found that the IEP for some symmetric NP shapes, namely, cubic and spherical NPs has the *f* values linear in a broad range of *n*. This allows to assume that the linearity of IEP in magnesium clusters can be connected to the more spherical shape of the most favorable structures. Indeed, it was established previously [18] that the $E_{rel}$ isomers of 1237 clusters of $Mg_2$-$Mg_{32}$ decreases linearly when the deformation parameter $DP=2P_3/(P_1+P_2)$ (where $P_1>P_2>P_3$ are the principal moments of inertia of the cluster) goes to 1 which means that the spherical structures are more energetically favorable. However, this tendency is quite fuzzy, with low regression coefficient and large deviations from the linearity as it is seen on Fig.11 of ref [18]. The Figure A2 of Appendix shows the plot of $\alpha_{iso}(n,m)$ values *versa* the *DP* of 1237 structures of $Mg_2$-$Mg_{32}$. It is seen, that there is no any mutual dependence of these parameters, not for most favorable isomers, nor for all clusters. Thus, the sphericity of isomers has no visible influence on IEP and cannot explain their features described above.

The polarizability of NPs is frequently described in the well-known induced dipole approximation introduced to chemical physics by Applequist. [32] This approximation proposes that the external electric field $\mathbf{E}_0$ induces the dipoles $\mathbf{p}_j$ ($j$=1,2,…$n$) located on the atomic centers of NP, and the induced atomic dipoles induce the additional field which modifies the dipoles of other atoms. The combined external and dipole field on the center $i$ ($i$=1,2,…$n$) is described by the formula:

$$\mathbf{E}_i = \mathbf{E}^0 - \sum_{j \neq i} \mathbf{T}_{ij} \mathbf{p}_j \qquad (3)$$

Here, $\mathbf{T}_{ij}$ is the 3x3 matrix describing the interaction of dipole moments of each pair of atoms $i$ and $j$ with the Cartesian coordinates $x,y,z$ and separated by the distance $r_{\mathrm{ij}}$:

$$\mathbf{T}_{ij} = \frac{f_e}{r_{ij}^3}\mathbf{I} - \frac{3f_t}{r_{ij}^5}\begin{pmatrix} x_i x_j & x_i y_j & x_i z_j \\ y_i x_j & y_i y_j & y_i z_j \\ z_i x_j & z_i y_j & z_i z_j \end{pmatrix} \qquad (4)$$

In the case of point dipoles (frequently referred as point dipole model of Applequist), the coefficients $f_{\mathrm{e}} = f_{\mathrm{t}} = 1$. Combining the above formulas, the self-consistent induced atomic dipoles can be expressed as

$$\mathbf{Bp} = \mathbf{E}^0; \quad \mathbf{B}_{ij} = \begin{cases} \alpha^{-1}, i = j \\ \mathbf{T}_{ij}, i \neq j \end{cases}; \quad \mathbf{p} = \mathbf{B}^{-1}\mathbf{E}^0 . \qquad (5)$$

The complete polarizability of a structure is $\mathbf{P} = \sum_i \mathbf{p}_i$ , and, thus, the polarizability of the system of the atomic dipoles is expressed with

$$\mathbf{P} = \boldsymbol{\alpha}\mathbf{E}_0; \quad \boldsymbol{\alpha} = \sum_{i,j} [\mathbf{B}^{-1}]_{ij} . \qquad (6)$$

It was well-recognized [33,34] that the description based on the formulas (3)-(6) suffer on so-called “polarization catastrophe” when the polarizability become infinite at some atomic arrangements. The solution of this problem was given for the first time by Thole [33,34] who proposed to replace the point dipoles by some charge distributions smeared around the atomic centers. Following this idea, several distribution models were proposed including the Thole's linear and exponential charge distributions [33,34], Ren and Ponder (AMOEBA FF) model [35,36], or Gaussian smearing [37-41]. All these models can be described by the same expression (3) with different $f_e$ and $f_t$. (see e.g. [38] for details). In all these modified dipole models, two adjustable parameters are used: the charge smearing parameter $a_{smear}$ (coefficient present in the expressions for $f_e$ and $f_t$, dependent on the model in use), and the atomic polarizability α.

We applied five above-mentioned models for the description of cluster polarizability using the DFT-optimized structures and energies of clusters. Namely, the Applequist model (AM), Thole linear model (TL), Thole exponential model (TE), Ren and Ponder model (RP), and Gaussian-distributed dipoles (GD) were explored. In these models, instead of using the adjustable value of α, we used the fixed Mg atom polarizability calculated by DFT. Making adjustment of $a_{smear}$ during the fitting of the $\alpha_{iso}(n,m)$ values calculated by formulas (3) against the corresponding DFT values, we obtain rather good coincidence between the DFT values of IEP, $\alpha_{iso}$(DFT), and approximate values $\alpha_{iso}$ calculated in the point dipole model, $\alpha_{iso}$(AM), Thole's linear and exponential models $\alpha_{iso}$(TL) and $\alpha_{iso}$(TE), Ren and Ponder model $\alpha_{iso}$(RP) and model of Gaussian dipoles $\alpha_{iso}$(GD).

The regression coefficients of $\alpha_{iso}(n,1)$ and corresponding statistics obtained with these models are shown in Table 1 (columns "Induced dipole models"). As is evident from the Table, the point dipole model of Applequist only poorly describes the DFT values both due to large final deviations between DFT results and fitted values, and due to poor linear dependence of $\alpha_{iso}(n,1)$. They also give the negative values of $\alpha_{iso}$ in many cases which is a non-physical result. At the same time, all the models of smeared dipoles well reproduce the DFT results, with best results achieved for RP and GD models, without any negative values of IEP. The dependence of $\alpha_{iso}$ on $n$ and $m$ estimated in these models are shown in Fig.6.

(*a*)

(*b*)

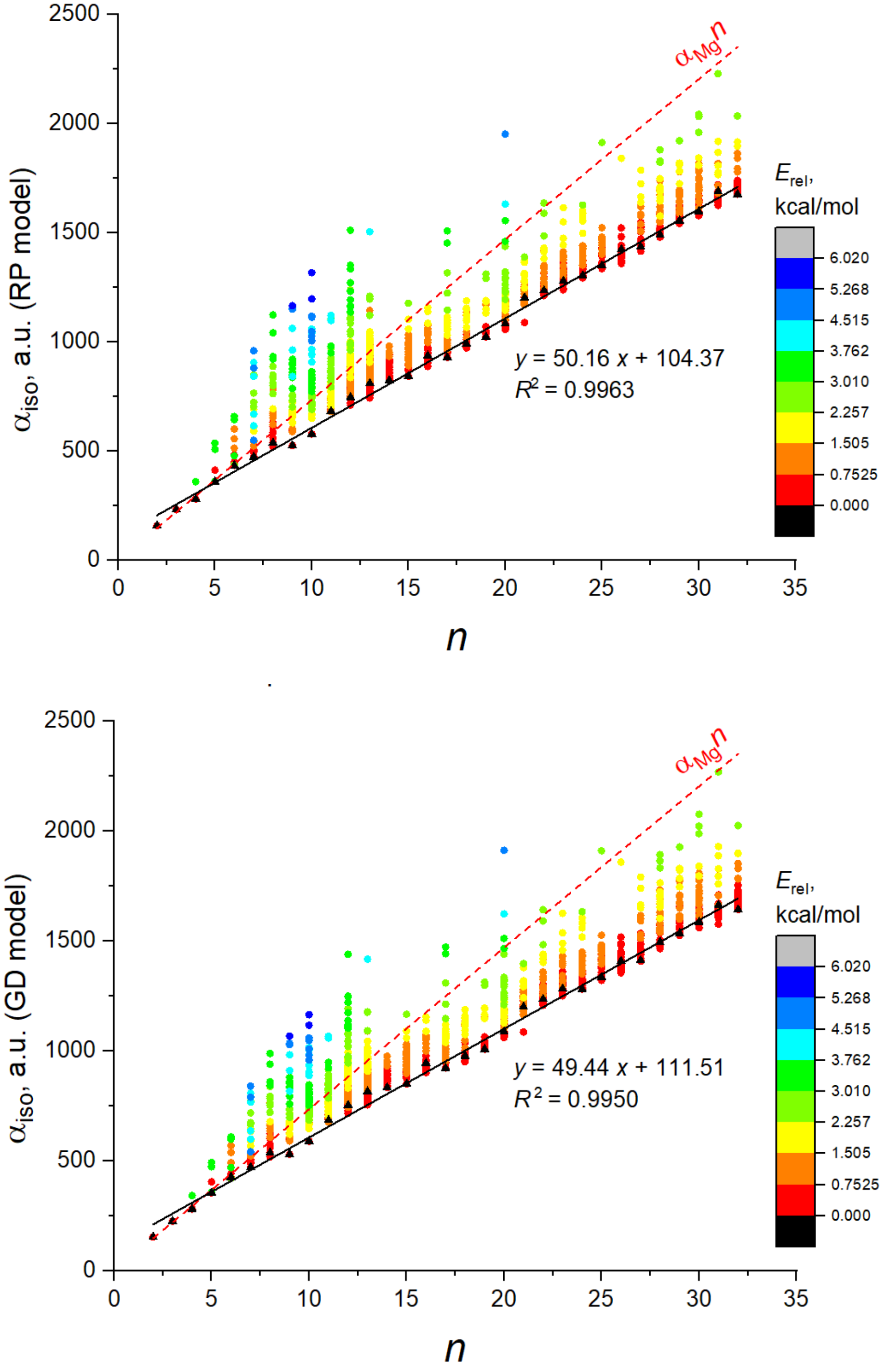

Figure 6. IEP of 1237 isomers of $Mg_2$-$Mg_{32}$ estimated in two models of induced dipoles: (*a*) RP model; (*b*) GD model. Color designates the relative energies of clusters calculated at the DFT level.

Among these models, the best fitted RMSD value for all 1237 IEP datapoints are achieved for the RP model. All models reproduce well the linearity of $\alpha_{iso}(n,1)$, as it can be concluded from the corresponding values of $R^2$, *SSE* and *SE*. It is interesting that the linearity of $\alpha_{iso}(n,1)$ is even better reproduced by some models than it takes place at the DFT level. The highest $\alpha_{iso}(n,1)$ linearity is achieved by the TE model although both Thole's models give rather poor RMSD values for the complete set of isomers. From this point view, the RP and GD models give better IEP estimates for the magnesium clusters. It is also worthwhile to investigate how these models reproduce the feature of minimality of $\alpha_{iso}$. This comparison is shown in Fig.7 where two parameters are analyzed: the number of "improper" values of $\alpha_{iso}(n,m)$ (i.e. the values which are less than $\alpha_{iso}(n,1)$) and the $E_{rel}$ energy interval where the clusters with "improper" $\alpha_{iso}$ are situated. This interval shows how important the deviation from the minimality property in these models. Analyzing the data on Fig.7, we conclude that the GD model is somewhat better reproduce these properties of DFT, because its patterns of both analyzed values are more similar to DFT than it takes place for RP and Tholes' models. It is also worthwhile to mention that the GD model, like in the case of DFT, gives the small range of $E_{\mathrm{rel}}$ values (less than ~1 kcal/mol) for all improper clusters except $n$=6 which allows concluding that the PMT principle is reproduces with a good accuracy.

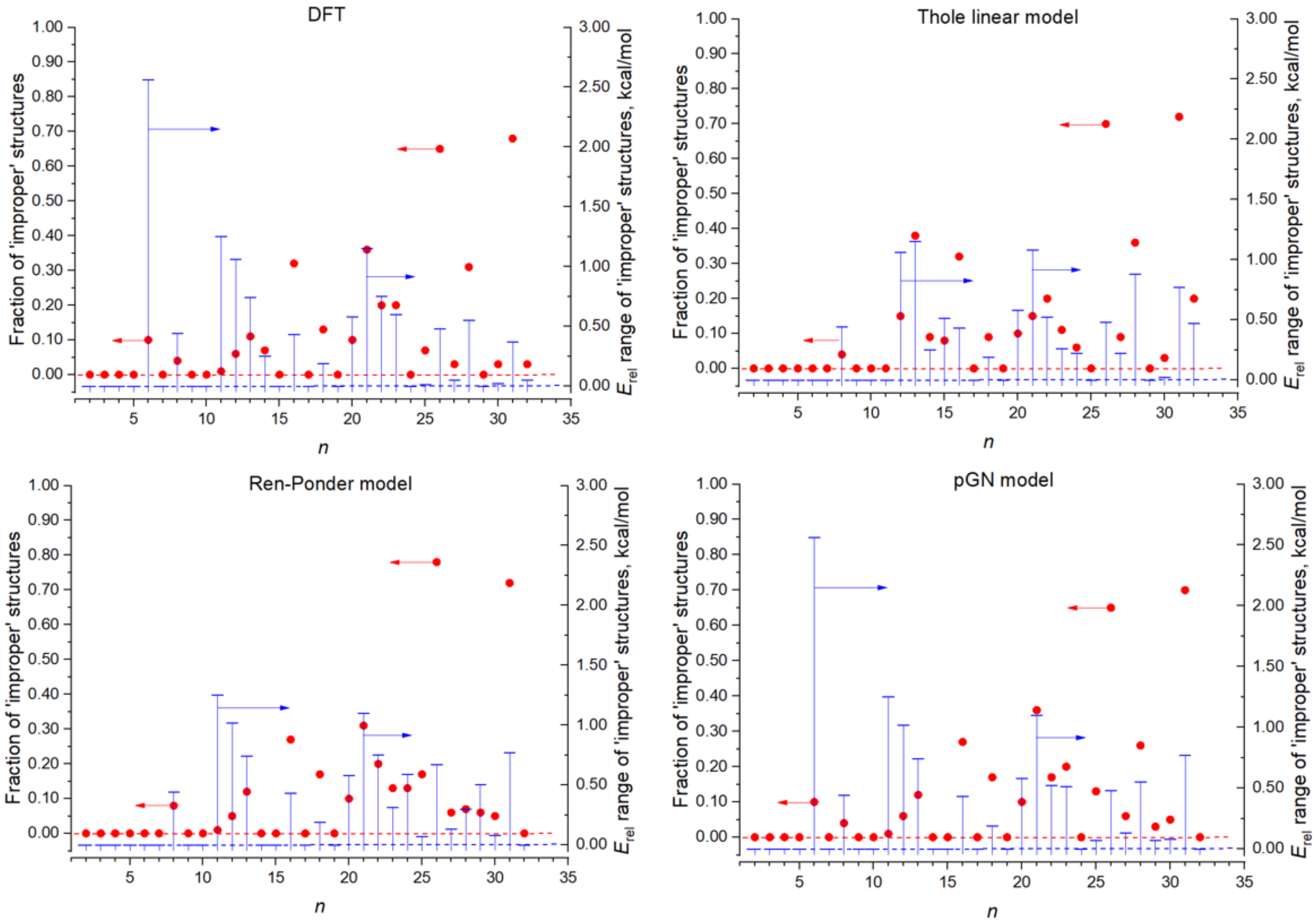

Figure 7. Deviations of the isotropic polarizability $\alpha_{iso}$ of most favorable $Mg_2$-$Mg_{32}$ structures from the minimum $\alpha_{iso}$ among the structures of the given nuclearity *n*. Left axis (red circles) – number of "improper" structures (i.e. structures with $\alpha_{iso}$ less than $\alpha_{iso}$ of the energetically most favorable structure); Right axis (blue bars) – relative energy range where the improper structures situated (energies of improper structures relatively to the most favorable one). Four panels show the results of direct DFT calculations along with three models for simple polarizability estimation.

The success of description of $\alpha_{iso}$ with such simple models encourages describing the PMT principle on this basis. However, the strong obstacle on this way is that the $E_{rel}$ values analyzed on Fig.7 are the DFT calculated values. All the attempts to reproduce the minimality feature of IEP using the energies calculated on the basis of dipole models were unsuccessful. Namely, Fig.A3 of Appendix demonstrates comparison between the isomers' energies calculated by DFT and their electrostatic energy components within the GD model for 1237 cluster isomers $Mg_2$-$Mg_{32}$. The Figure shows the various combinations of monopole-monopole $U_{00}$, monopole-dipole $U_{01}$, dipole-dipole $U_{11}$ energy contributions, and the self-energy of induced dipoles $U_{ss}$ calculated by the formulas of GD model as reported in ref [40]. Figure demonstrates lack of correlation between the isomers' relative energies $E_{rel}$ calculated by DFT and $E_{rel}$ estimated on the basis of any GD model energy components or their combinations. In the case of perfect correlation, we expect the linear dependence close to the bisector of the first quadrant. However, as is seen from the plots, the calculated energy patterns are absolutely different from the energy dependencies obtained in the DFT calculations. We failed to find any combinations of electrostatic energies obtained within the induced dipole model as well as with additional energy contributions of the interacting atomic point charges (both nuclear and DDEC6 derived) which could reproduce the energy dependence similar to the DFT energies. Thus, the dipole models, although reproduce the linearity property, do not reproduce the property of minimality. Moreover, reproducing the linearity of clusters IEP, these models have no direct "explanation" of this feature. It is not clear why some structure has such a linear dependence of $\alpha_{iso}$ whereas other ones do not. Obviously that these features are connected to any deeper regularities in the Kohn-Sham (or Schrödinger) equations describing this chemical system, and further studies are needed to elucidate these regularities.

**Concluding remarks**

We calculated the electrostatic polarizability of large set of sub-nanoscale magnesium clusters comprising 1237 structurally unique isomers. The polarizability calculation reveals that the IEP of the most favorable isomers on the cluster nuclearity *n* is linear with a high correlation coefficient, and its value for each *n* is close to the minimum value among all found isomers of a given nuclearity. These features are quite unusual, given that the cluster structures that exhibit these properties are very different. This suggests that the favorability of the cluster structure is closely related to their polarizability and, possibly, the atoms forming the cluster tend to arrange themselves in such a way as to provide the most dense or compact packing of the electron density. At present, it is not clear how general this property is, and additional studies are required to establish whether this property is fulfilled in clusters of other metals, as well as in clusters of a more complex elemental composition. If it turns out that these properties manifest themselves in a wide range of systems, this will open up opportunities for developing new methods for predicting the structure of polyatomic nanostructures and searching for global minima of sub-nanoparticles, an issue that currently causes considerable difficulty in solving. Additional studies are also required to clarify the nature of the observed patterns. We have shown that the property of linearity is reproduced at the level of simple induced dipole models. However, these facts do not allow us to reveal the true reason for the property of energy minimality of the least polarizable systems. We should also note that, at the moment, the dipole approximation does not elucidate the origin of linearity directly, it only reproduces it. Thus, the nature of the established features remains intriguing question of modern chemical physics.

**Acknowledgements**

S.K.I. acknowledges the support of Russian Foundation for Basic Research (project No. 20-03-00282). Authors thanks Ilya S. Steshin for his assistance in the visualization of the cluster structures carried out with his program AgloView.

# APPENDIX

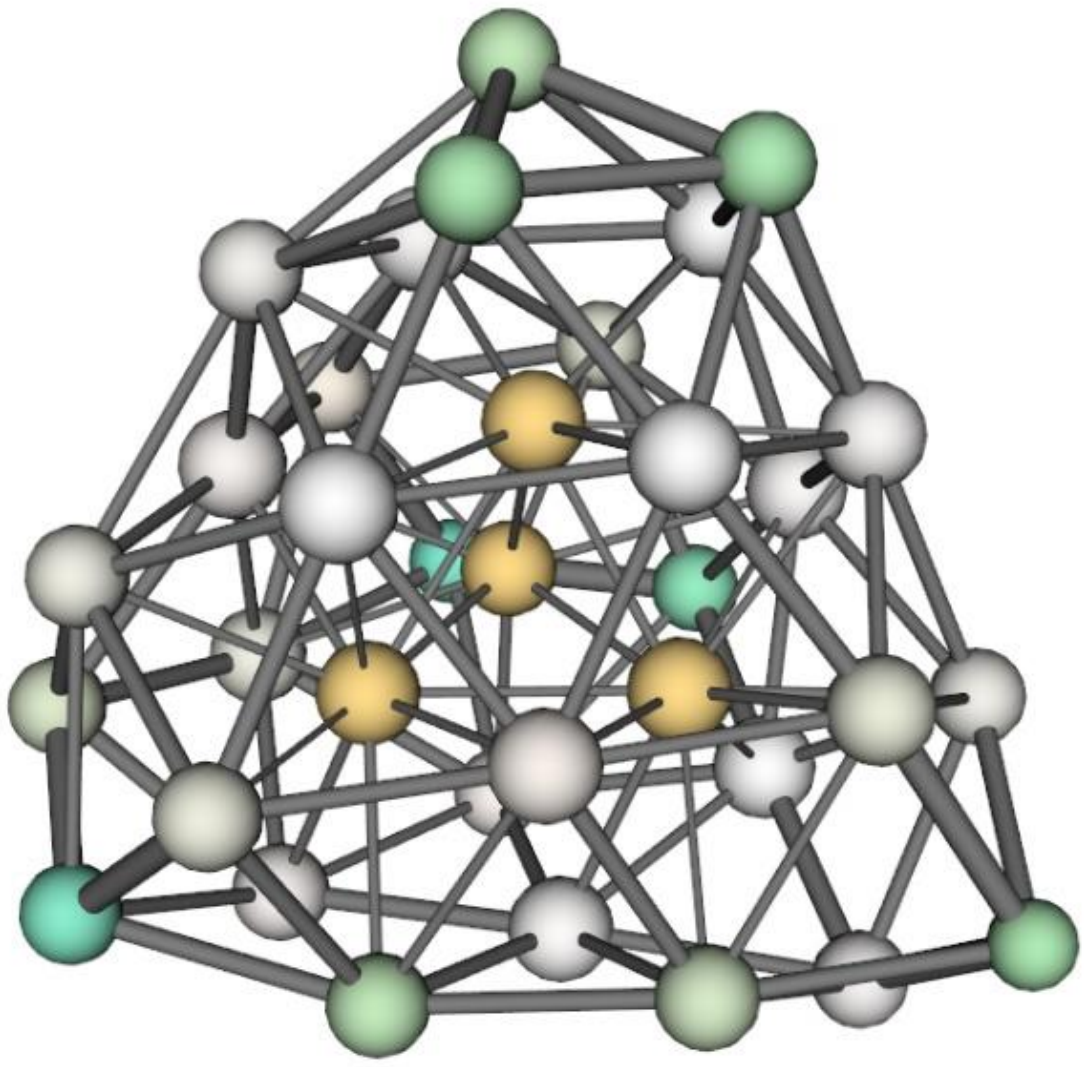

Figure A1. Global minimum structure of cluster $Mg_{35}$ predicted by the MTP potential

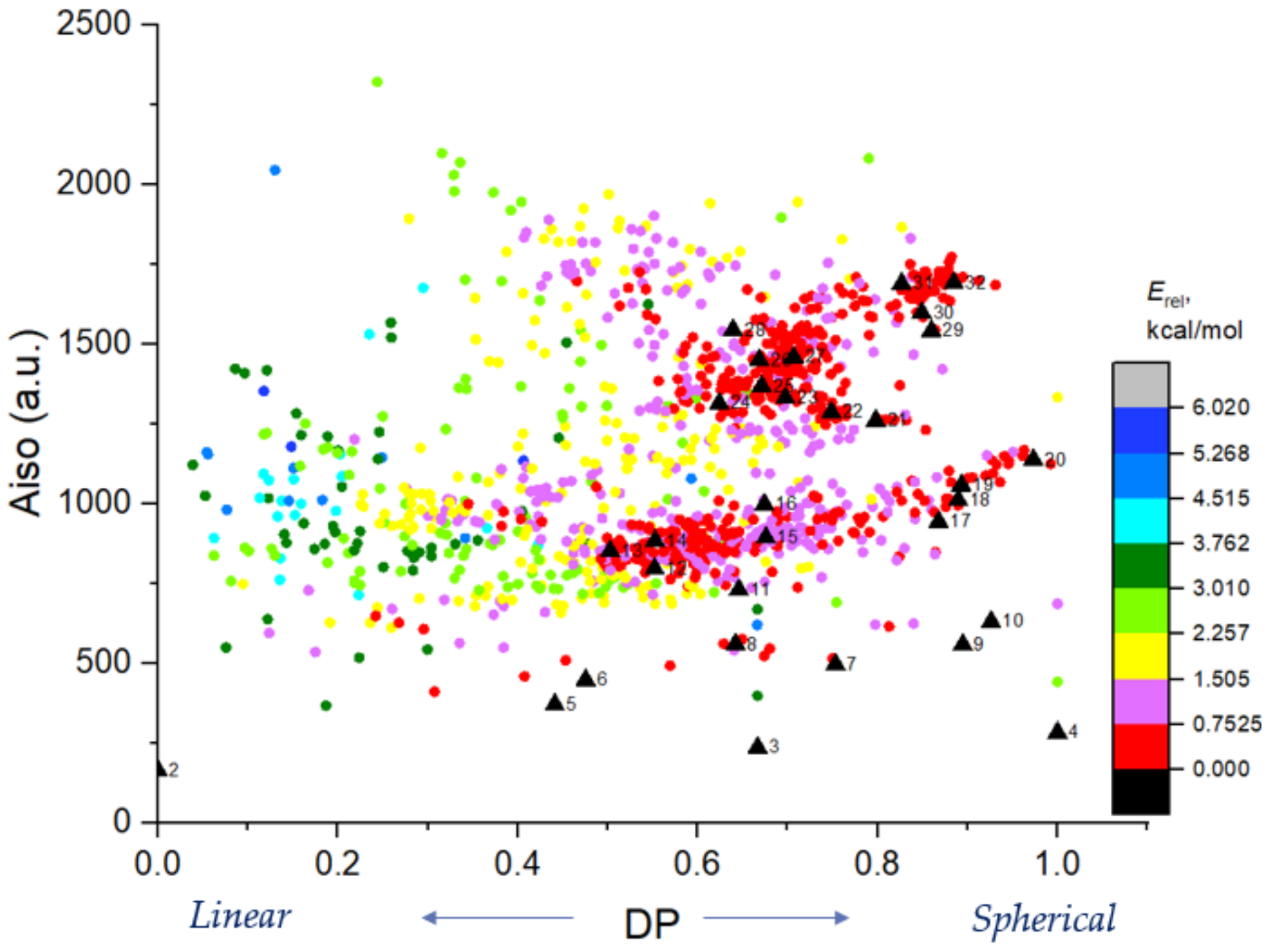

Figure A2. Dependence of DFT-calculated IEP of 1237 isomers of $Mg_2$-$Mg_{32}$ on the structure deformation parameter *DP*. Color designates the isomer relative energy. Black triangles correspond to the most favorable structures of nuclearity indicated by numbers.

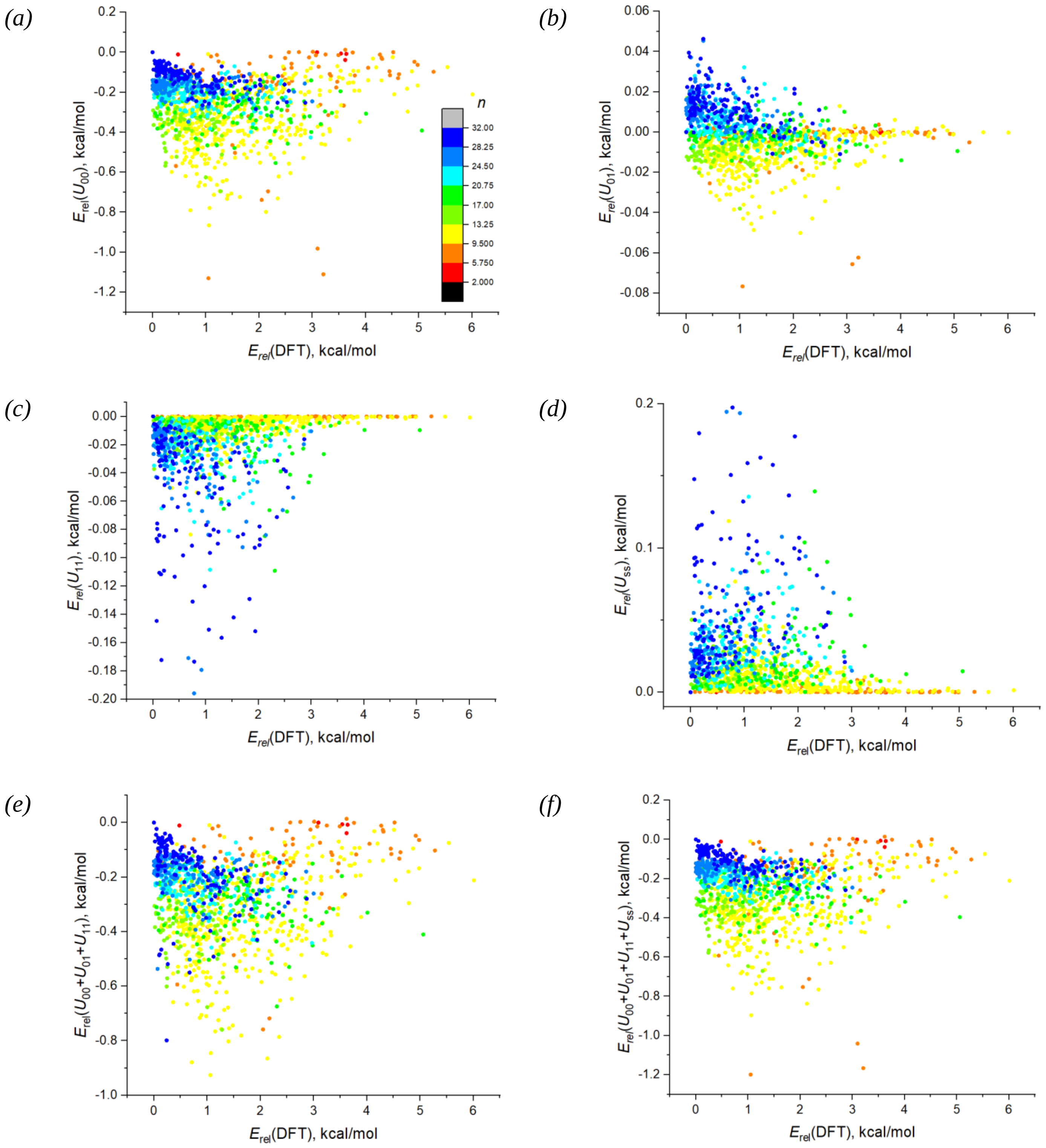

Figure A3. Lack of correlation between the isomers' relative energies $E_{rel}$ calculated by DFT and estimated on the basis of various energy components within the GD model ($U_{00}$, $U_{01}$, $U_{11}$, $U_{ss}$ and their combinations) for 1237 cluster isomers $Mg_2$-$Mg_{32}$. Color indicates the cluster nuclearity (single scale for all charts). Some data points outside the vertical axis ranges are omitted.